\documentclass{article}
\usepackage{amsmath}
\usepackage{amsthm}
\usepackage{amsfonts}
\usepackage[final]{graphicx}
\usepackage{makeidx}
\usepackage{epsfig}

\begin{document}
\title{A Cubic Whitney and Further Developments in Geometric Discretisation}
\author{
Samik Sen\footnote{e-mail:samiksen@hotmail.com},\\ 
{\em  School of Mathematics, University of Dublin, Dublin 2, Ireland.} 
}
\date{}
\maketitle
\begin{abstract}
Geometric discretisation draws analogies between discrete
objects and operations on a complex with continuum ones on a manifold.
We generalise the theory to the cubic case and incorporate
metric, by adding volume factors to our discrete Hodge star and
then by modifying our inner product which leads to the same result.
\end{abstract}

\section{Introduction}
Geometric discretisation (GD) \cite{Adams, Us} appears quite complicated, at first,
using unfamiliar maps and objects but nothing could be further from the truth.
Simply put we can translate from continuum objects to discrete ones, and back again
in such away that not only can we make the discrete ones as close as we want to the
continuum ones, but we have discrete operators which satisfy the same 
identities as
their continuum counterparts. We have $d^2=0$, Stokes' theorem, the Leibniz rule
and more; meaning that the discrete theory mimics the continuum one to a
remarkable extent. Here we are interested in the theory itself, as opposed to 
applications of which there are many. In particular, we introduce a cubic Whitney map
\cite{me} and metric, needed for the discrete Hodge star which is after all not purely a topological
object.

The basic structure of the theory, using the Whitney and de Rham map, to translate
from the discrete the the continuum and back again, is the same as considered by
Dodziuk \cite{Dodziuk}. GD uses a subdivided space though, in order to have a discrete
Hodge \cite{Adams}; Hiptmair \cite{Hiptmair} considers the discrete Hodge star using finite
elements which map also to dual spaces and satisfy $\delta=*d*$, where he refers
to cubic work done by Nedelec \cite{Nedelec} which I have only recently come across.
My focus has been on application to lattice field theory \cite{fermions}, though
the relationship with finite elements has also been of interest. 

GD deals with operators, as well as functions, in such a way that the identies and rules which they obey 
hold. Notably we have
\begin{itemize}
\item Stokes' Theorem (Gauss' Law in electromagnetism)
\item The Leibniz rule (The product law for differential forms)
\item A discrete Hodge star $\star$ where $\star^2=I$ and $\delta=\star d \star$.
\item $d^2=0$
\item A skew symmetric wedge
\end{itemize}
and DO NOT have associativity of the wedge. The presence of a Hodge star with
the associated properties is the most significant feature of the system.

There are two aspects to our work here. First, we introduce a cubic Whitney map
and then tackle convergence. We find that by the addition of 
volume factors to the Hodge star we can demonstrate convergence. This is not
entirely satisfactory since the inner product is intimately related to $\star$; 
we cannot alter one without the other. That being the case, we look
at how a natural inner product leads to the introduction of exactly the volume
factors which we had anticipated above.

We have also been considering the implications of this theory to lattice field
theory, where the lack of a discrete Hodge star has been a problem \cite{Rabin} 
but have been aware that related work has been ongoing from engineering 
\cite{Hiptmair, Nedelec, Bossavit}. The possibility of fruitful cross-fertilisation is 
very much on the cards and something which we have always been interested in developing further. 

In short we begin with a review of GD before going on to our new work where:
\begin{enumerate}
\item We develop a cubic theory. 
\item We demonstrate converge using a heuristic involving 
``natural volume factors''.
\item We show that convergence, using a new inner product which retains the
relationship of the Hodge star to the inner product, leads to 
precisely the same factors as we used for the heuristic.
\end{enumerate}
We finish with a brief discussion of our current and future work.

\section{Geometric discretisation}
%
Geometric discretisation (GD) \cite{Adams} is a discretisation scheme 
based on a correspondence between discrete objects and operations on
a complex, $K$, \cite{Nakahara, NS} with continuum ones on a manifold, $M$, \cite{Eguchi} 
which captures topology \cite{Us}.

The de Rham, $A^K$, and Whitney map, $W^K$, play a central role 
since they allow us to move from continuum to discrete and back again
 \cite{Dodziuk} whilst maintaining topology\footnote{Both these operators 
commute with $d$, so if $f$ is exact or closed in the continuum, it is also in 
the discrete.}. For the moment, we note that this provides a very clean 
structure and allows us, for example, to induce a natural discrete 
wedge from the continuum one
\[ x \wedge^K y = A^K(W^K(x) \wedge^K W^K(y)),\]
which inherits skewsymmetry and the Leibniz rule; though not associativity.

The other key idea is that of a subdivided space $B$, containing both 
the original, $K$, and dual, $L$, spaces, which allows us to introduce a
discrete Hodge star operator \cite{Adams}. This has the property that $\star\star=1$
and $\delta=\star d\star$, with appropriate signs, which is of interest \cite{Rabin}.
For example, we are able to capture chirality in the Dirac-K\"ahler 
formalism as a result \cite{fermions}.

In summary we have the following in GD, with the noticeable exception 
of associativity for the wedge:
\begin{itemize}
\item An exterior derivative which maps from $p$-cochains to $(p+1)$-cochains with $d^2=0$.
\item A wedge product, with which we can take the product of a $p$-cochain
and a $q$-cochain to get a $(p+q)$-cochain. This has the properties that
\begin{itemize}
\item Skewsymmetry:$x^p \wedge y^q = (-1)^{pq} y^q \wedge x^p$.
\item Leibniz Rule:$d(x^p \wedge y^q)= dx^p \wedge y^q + (-1)^p x^p \wedge (dy^q).$
\end{itemize}
\item The Hodge Star: This duality map associates an $(n-p)$-cochain in $L$ to each 
$p$-cochain in $K$, capturing $\star\star=I$ and $\delta=\star d\star$ with appropriate signs.
\end{itemize}

For associativity we get
\begin{equation}
(x^p\wedge^K y^q) \wedge^K z^r = \left(\frac{p+q+1}{r+p+1}\right) x^p\wedge^K (y^q \wedge^K z^r).
\label{equation:nawedge}
\end{equation}

Given a triangulation\footnote{We do not worry about how to construct a 
triangulation of the 
space we are working with though we do know that they exist for the cases we 
are interested in since Rad\'o\cite{Rado} proved this for compact spaces.
In fact we know any differentiable manifold can be triangulated\cite{HY}.}
of our space we can translate from differential geometry to our discrete
structure. We start by looking at how forms are projected onto the triangulation
and back before moving on to various operations.
\begin{figure}[h]
\begin{center}
\includegraphics[width=150pt]{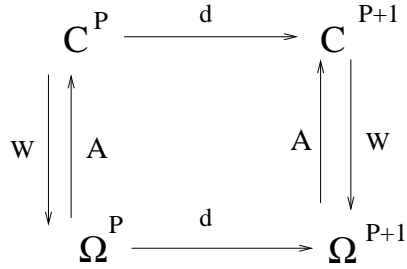}
\caption{$d$, $A^K$ and $W^K$ provide a commutative diagram if we restrict ourselves to 
the image of the Whitney map (a.k.a. Whitney elements), a finite dimensional space of functions.}
\label{fig:comm_diag}
\end{center}
\end{figure}

\noindent
\subsection{de Rham map}
$p$-cochains, $C^p(K)$, on the triangulation are the discrete analogies of $p$-forms, 
$\Omega^p(M)$, on the manifold. The $p$-forms can be projected onto the 
$p$-cochains using the de Rham map, $A^K$,  which involves integrating them over 
the associated $p$-chains. In other words we have $A^K:\Omega^p(M) \to
C^p(K)$ defined as
\[ <A^K\omega^p,\sigma^p_i> = \int_{\sigma^p_i} \omega^p,\]
which has the property that $dA^K=A^K d^K$ (Stokes' Theorem).

\begin{figure}[h]
\begin{center}
\includegraphics[width=100pt]{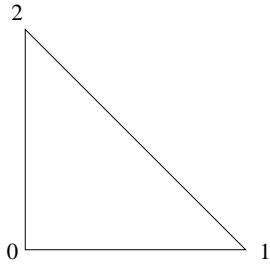}
\caption{The standard triangle.}
\label{fig:stdtri}
\end{center}
\end{figure}

If we take the standard triangle $[0,1,2]$, shown in Fig.\ref{fig:stdtri}, 
we can see that what happens explicitly. Since this is in $2$D we have $1, dx, dy$ 
and $dxdy$ as our
possible differential forms, all with possible function coefficients $f$. The possible 
cochains are $[0], [1], [2], [0,1], [0,2], [1,2]$
and $[0,1,2]$. $1$ is a $0$-form and so is mapped onto the vertices. This means that
$f$ is mapped to $f([0])+f([1])+f([2])$. Similarly $fdx$ is mapped to 
$(\int_{[0,1]} fdx) [0,1] + (\int_{[0,2]} fdx) [0,2]+ (\int_{[1,2]} fdx) [1,2]$. 

\subsection{Whitney map}
The Whitney map is the complimentary operation, from $p$-cochains to $p$-forms. 
In order to introduce this we need the barycentric coordinates, $\mu_i$'s. 
Given an $n$-dimensional complex we have $\mu_0,\dots,\mu_n$ defined on each
$n$-simplex which have the property that:
\begin{itemize}
\item $\mu_i([v_j])=\delta_{ij}$.
\item $\sum_i \mu_i(x)=1$ for all $x$ inside the triangle.
\item $\mu_i=0$ outside the triangle.
\end{itemize}
In our standard triangle, with vertices with coordinates $(0,0)$, $(1,0)$ and $(0,1)$,
we can define 
\begin{eqnarray*}
\mu_0 &=& 1-x-y \\
\mu_1 &=& x \\
\mu_2 &=& y 
\end{eqnarray*}
which satisfy the conditions necessary.

We can then define $W^K$ as 
\[ 
W^K[v_0,\dots,v_p]=\sum_i p!(-1)^i \mu_i d\mu_0\wedge \cdots  \hat{d\mu^i} \cdots 
\wedge d\mu_i
\]
where $\hat{d\mu^i}$ denotes that $d\mu_i$ is emitted.
With this we have that $dW=Wd$ \cite{Dodziuk, Whitney} and $A^KW^K=I$.
This leads to a commutative diagram of sorts, Fig. \ref{fig:comm_diag},  since we can starting from a 
form, map it to the triangulation and then act on it with $d$ or act on it
with $d$ before taking our approximation; either way we get the same result.
It is a ``commutative diagram'' since $W^KA^K$ is not equal to the identity, which means
that not all routes are equivalent. 

\subsection{Wedge}
Having established the basic mechanic of using $W^K$ and $A^K$, to map to and fro between
our spaces, we can induce a discrete wedge product $\wedge^K$ on the discrete side: 
\[ x^p \wedge^K y^q = A^K(W^K(x^p)\wedge W^K(y^q)).\]
This is both distributive and anti-symmetric but not associative\footnote{It is an
anti-symmetrisation like \cite{starproduct} and \cite{Albeverio}}, as can be seen 
from Eqn.\ref{equation:nawedge}. 

\subsection{Hodge star}
\begin{figure}
\begin{center}
\includegraphics[width=150pt]{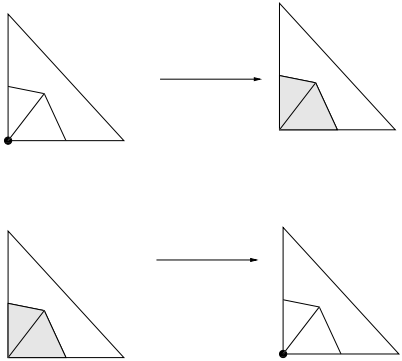}
\caption{A dual space leads to $\star\star=I$ type behaviour.}
\label{fig:star1}
\end{center}
\end{figure}

\begin{figure}
\begin{center}
\includegraphics[width=150pt]{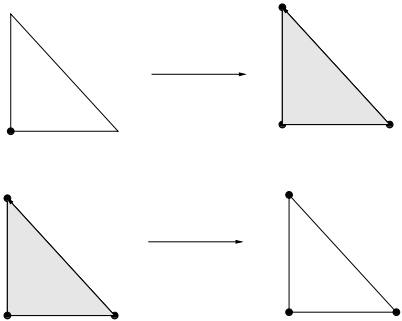}
\caption{Without a dual space $\star$ doesn't know which vertex to map back to and so end up
mapping to all of them.}
\label{fig:star2}
\end{center}
\end{figure}

\begin{figure}
\begin{center}
\includegraphics[width=150pt]{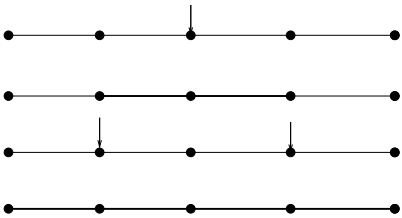}
\caption{Without a subdivided space we do not get $\d=\star \delta \star$ type behaviour. Initially $\star$
maps the vertex to the edge, before $\partial$ maps to its edges. Finally the second $\star$ maps these
vertices to edges which results in a line which is twice as long as it would have been if it initial vertex
had been acted on with $d$. The use of a subdivided space halves the lengths and thereby fixes the problem. 
Of course multiply both sides by $\star$ leads to $d=\star \delta \star$.}
\label{fig:star3}
\end{center}
\end{figure}

The Hodge star is the jewel in the GD\footnote{This was developed by
Adams \cite{Adams}} crown. 
There are various problems with having a discrete star
satisfying both $\star\star=I$ and $\delta=\star d\star$ type behaviour
as discussed by Rabin \cite{Rabin}.
These can be resolved by working in a subdivided space and saying that
$\star$ maps from the original triangulation to a dual space. The advantage
of this can be seen from the Figures \ref{fig:star1}-\ref{fig:star3}.

First from Fig. \ref{fig:star1} we see that with a dual space we have a 
trivial identification of original and dual objects and so capture $\star\star=I$; while 
in Fig. \ref{fig:star2} we see that without a dual space, we do not
return to where we began and so do not have $\star\star=I$.
Finally, in Fig. \ref{fig:star3}
we see that without the use of a subdivided space we do not
get $\delta=\star d\star$ but end up linking vertices to vertices ``two units away''.

Aside from associativity of the wedge, we lack a Whitney map on the dual
space. Partially motivated by this, we develop a cubic version of the theory next, 
whose dual space is also cubic.

\section{Cubic theory}

%
%
We begin by introducing our notation before looking at 
the operators $d$ and $\partial$ in this language and moving on to the de 
Rham and Whitney maps, which we show to have the desired properties.

Note that the dual to a cube is also a cube which means that we also have a 
Whitney map from the dual space which was not the case with simplices.
We unfortunately were unable to develop a generic method to determine Whitney maps, 
which is the tricky part to generalise, for arbitrary cells.

\begin{figure}
\begin{center}
\includegraphics[width=100pt]{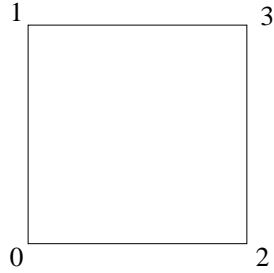}
\caption{Binary notation applied to a square, where vertex number expressed in binary leads
leads to the $x$ and $y$ coordinates of that vertex. This makes it very easy to know which combinations
of vertices are edges since they must differ by a power of two. In 3D for example we can tell that the
edge [2,3] is connected to [6,7].}
\label{fig:square1}
\end{center}
\end{figure}

\subsection{Cubic Notation}
We describe the various vertices, edges and faces, taken from  Fig. \ref{fig:square1}, in our new notation below:
\begin{center}
\begin{tabular}{lc}
Vertex 0 & $[0,0]$ \\
Vertex 1 & $[0,1]$ \\
Vertex 2 & $[1,0]$ \\
Vertex 3 & $[1,1]$ \\
Edge from 0 to 1  & $[0,y]$ \\
Edge from 0 to 2  & $[x,0]$ \\
Edge from 1 to 3  & $[x,1]$ \\
Edge from 2 to 3  & $[1,y]$ \\
Face ``[0,1,2,3]'' & $[x,y]$\\
\end{tabular}
\end{center}
In n-dimensions, our general object, for the unit hypercube, is thus $[a_0,\dots,a_n]$, where
$a_i$ can be 0, 1 or $x_i$.

\subsection{Cubic Operations}
We define $D=d_{cube}$ as follows:
\begin{equation}
 D[a_0,\dots,a_n]=\sum_i s_i p_i[a_0,\dots,a_n], 
\label{equation:D}
\end{equation}
where $s_i=0$ if $a_i=x_i$, $s_i=+1$ if $a_i=1$ and $s_i=-1$ if $a_i=0$. $p_i$
equals $(-1)^n$, where $n$ is the number of $a_j$'s equal to $x_j$ for $j<i$.

Next we introduce the boundary operator:
\[ \partial:[a_0,\cdots,a_n]=\sum_i p_i ([a_0,\cdots,1,\cdots,a_n]
-[a_0,\cdots,0,\cdots,a_n]),\]
where we have replaced the ith slot with a $0$ and a $1$, the second one
having opposite the orientation. $p_i$ is the same as for $D$.
In other words
\[ \partial[x]=[1]-[0]\]
and
\[ \partial[x,y]=-[0,y]+[1,y]+[x,0]-[x,1]\]
which is what we expect the boundary operation to be. To test it more 
thoroughly we apply it to $[x,y,z]$:
\begin{align*}
\partial [x,y,z] &= -[0,y,z]+[1,y,z]+[x,0,z]-[x,1,z]-[x,y,0]+[x,y,1] \\
\partial^2 [x,y,z] &= [0,0,z]-[0,1,z]+[0,y,1]-[0,y,0]+[1,1,z]-[1,0,z]+ \\
&[1,y,0]-[1,y,1]+[1,0,z]-[0,0,z]+[x,0,0]-[x,0,1]-[1,1,z]+[0,1,z]+\\
&[x,1,1]-[x,1,0]-[1,y,0]+[0,y,0]+[x,1,0]-[x,0,0]+[1,y,1]-[0,y,1]-\\
&[x,1,1]+[x,0,1] \\
&=0.
\end{align*}
The boundary map is clearly taking us to faces of the object it is acting
on but from this we can see that the orientation of the various factors is 
being dealt with properly too.

We want to show that $\partial^2=0$ in general. We can see that 
$\partial$ maps an $m$-cell to two oppositely oriented pieces. We say that 
the first $\partial$ sets $a_k$ to either $0$ or $1$ and the second the same to $a_l$. 
We denote this as
\[ \partial \sigma^m = \sigma^{m-1}_0-\sigma^{m-1}_1 \]
in short hand, where we mean sum over $i$ where $a_i=0$ by $\sigma^{m-1}_0$.
Then 
\[ \partial^2 \sigma^m=
\sigma^{m-2}_{00}
-\sigma^{m-2}_{01}
-\sigma^{m-2}_{10}
+\sigma^{m-2}_{11}.\]
In fact we have two cases to consider depending on 
whether $k<l$ or $k>l$. The idea is that if $k<l$ then
we get the same sign factor whether the first $\partial$ removes $k$ or the 
second. This is not the case with $l$ since the $p_i$ factor changes.
Thus we get two oppositely oriented versions of the same term, which cancel.
The same argument applies if $k>l$. 

\subsection{de Rham and Whitney}
We next introduce $A^K$ and $W^K$ in the cubic framework, showing they satisfy
the desired properties
\begin{itemize}
\item $A^KW^K=I$
\item $dW^K=W^KD$
\item $DA^K=A^Kd$
\end{itemize}

The Whitney map is defined as
\begin{equation}
 W^K[a_0,\dots,a_n]=W^K[a_0]W^K[a_1]\dots W^K[a_n],
\label{equation:Whitney}
\end{equation}
where, for edges $[0,x_i]$ of length $h_j$,  we have 
\begin{align}
W^K[0] &=\frac{h_j-x_i}{h_j}, \\
W^K[1] &= \frac{x_i}{h_j}, \\
W^K[x_i] &=\frac{dx_i}{h_j}.
\end{align}

From Eqn.\ref{equation:Whitney}, the general case consists of products of terms
of the form $W^K[a_i]$. $W^K[a_i]$ maps to $\frac{h_j-x_i}{h_j}=\frac{h_j}{h_j}=1$ where $a_i=x_i=0$, 
to $\frac{x_i}{h_j}=\frac{h_j}{h_j}=1$ where $a_i=x_i=h_j$ and since $W^K[x_i]=\frac{dx_i}{h_j}$, we get
\[ \int_0^{h_j} \frac{dx_i}{h_j}=1, \]
when $a_i=x_i$. Thus $A^K W^K=I$ in general.

From Eqn.\ref{equation:D} and Eqn.\ref{equation:Whitney} we see that 
\[ 
W^KD[a_0,\dots,a_n]=\sum_i s_i p_i W^K[a_0]\dots W^K[a_n]
\]
and
\begin{align}
dW^K[a_0,\dots,a_n] &=\sum_i W^K[a_0]\dots dW^K[a_i]\dots W^K[a_n] \\
&=\sum_i s_i p_i dW^K[a_i]W^K[a_0]\dots \hat{W^K[a_i]}\dots W^K[a_n],
\end{align}
Thus
\[
dW^K=W^KD.
\]

Finally, from Stokes' theorem we see that $A^Kd$ acting on a $p$-form
\[ A^Kd \omega^p=\sum_j \int_{\sigma^{p+1}_j} d\omega^p [\sigma^{p+1}_j]=
\sum_j\int_{\partial \sigma^{p+1}_j} \omega^p [\sigma^{p+1}_j]=
\sum_{i,j} \int_{I_{i,j} \sigma^{p}_i}\omega^p[\sigma^{p+1}_j],\]
is the same as
\[ \sum_{i,j} I_{i,j} \int_{\sigma^p_i} \omega^p[\sigma^{p+1}_j]=
\sum_i d \int_{\sigma^{p}_i} \omega^p [\sigma^{p}_i]=
dA^K \omega^p.\]

For this to hold in the cubic case we need $\partial$ and $D$ to be compatible,
as they are in the simplicial case via the incidence matrix.

Looking at $\partial$ we can see that the sign, of incidence matrix elements, is 
determined by whether we introduce a $0$ or a $1$. This is also the case in $D$.
We get a (-1) contribution if we introduce or remove a $0$. The remaining factor is
the $p_i$ one which also occurs  in both $\partial$ and $D$. So the incidence matrix
 \cite{HY} 
associated with $\partial$ induces a $D$ which is the same as the one we have that 
been using. Since $D$ is compatible $A^KD=dA^K$ follows.

Note that since $A^KW^K=I$ we also know that
\[ D^2=(A^KW^K)D^2=d^2(A^KW^K)=0.\]

Thus we have 
\begin{enumerate}
\item Introduced a cubic notation.
\item Defined appropriate boundary and coboundary operator.
\item Developed a cubic Whitney map.
\item Shown that these satisfy the desired properties.
\end{enumerate}

Note that if we map $dx$ onto the standard triangle then due to the diagonal edge, $[1,2]$, we introduce
a $dy$ component when mapping back using $W^K$. This does not happen in the cubic.

Next we use this formalism to incorporate metric into geometric discretisation
and show that image of the discrete Hodge star converges to the continuum one.
\section{Incorporating metric via heuristic}
We have treated GD primarily as a topological theory but $\star$ has metric dependence 
which must be considered also. Since we can only compare two similar objects,
we map continuum ones to the triangulation and back again, using $W^KA^K$, and note the
difference. So when looking at functions, or forms, 
we consider 
\[ f-W^KA^Kf,\]
and its dependence on the size of the discrete cells used.

In 1D, for example, we can see that via the de Rham map, $f(x)$ goes to $f(0)[0]+f(1)[1]$. 
$W$ then maps this to $f(0)+(f(1)-f(0))x$; a piecewise linear approximation. 
For $fdx$ we get $\int_{[01]} fdx$ on the complex which $W$ maps to $(\int_{[01]} fdx)d\mu_0$. 
Since $\mu_0$ goes from  $0$ to $1$ as $x$ goes from $0$ to $L$, the length of $[01]$, we get 
\[
  W^KA^K(fdx)=W\int_{[01]}fd\mu_0=\frac{\int_{[01]}fdx}{L}
\]
or the average value of $f$ along the edge $[01]$.

What we find is that the approximation made involves taking the average of the 
continuum object for forms and piecewise linear approximations for functions. 
The de Rham map takes the integral over a triangle, say, while the 
Whitney map results in dividing this by the volume so
\[ f d^px \to (\int_\sigma^p f d^px) \sigma^p \to \frac{\int_\sigma^p f d^px}{
{\rm Vol } \sigma^p} = f_{\rm average} d^px.\]

We can analyse $\star$ using this picture. If we start with a $p$-form, we integrate it over
a $p$-simplex. This leads to a $p$-cochain in our discrete structure. We can act on this 
with $\star$, which maps it to an $(n-p)$-cochain which we can be mapped back to continuum
space with Whitney(if we had a dual Whitney map that is, which we do for the cubic case).
The problem then is that the Whitney map divides by the ``wrong'' volume; we no longer
have an average:
\[ f d^px \to (\int_{\sigma^p} f d^px ) \sigma^p \to 
\frac{(\int_{\sigma^p} f d^px)}{{\rm Vol } \sigma^{n-p}} d^px \neq f_{\rm average} d^p x. \]

The idea then is to introduce volume factors to $\star$ which fixes this. We say that
\[ \star_{NEW} \sigma^p \to \frac{{\rm Vol } \sigma^{n-p}}{{\rm Vol } \sigma^p} \star \sigma^p.\]
Using this we have anticipated the problem with $W$ so that now, when we divide by 
the ``wrong'' volume factor, it simply cancels with what we have, leaving the 
correct term to get the average. This also has the property that when we look 
at $\star\star$ the factors cancel out, as they should.

\label{section:heuristic}

Once we have this, we can see that $\delta=\star d\star$ and the Laplacian should also converge. 
For the first case we want $W^K\star d\star A^K$ to converge to the continuum case. We can 
rewrite this as $W^K\star A^KW^Kd\star A^K$ since $A^KW^K=I$, but this is $(W^K\star A^K) d (W^K\star A^K)$ since $dW^K=W^Kd$.
So if we have $W^K\star A^K$ converging to $\star$, which we do with the addition of the 
volume factors, we have the desired result. The Laplacian 
follows once we have $\delta$ converging since we can move $d$ through $A^K$ and $W^K$.
Note that we only have these result for the cubic case since we do not have
$\star$ acting on the dual space otherwise. Dodziuk \cite{Dodziuk} has previously 
shown convergence of functions, forms and various operators though his system was 
did not involve a subdivided space.

\section{Incorporating metric via inner product}
The previous modification is nice, since with it we have our discrete $\star$ converging 
to the continuum one BUT it is against the spirit of GD. By adding an {\it ad hoc} term 
we break the relationship between the inner product and $\star$.
We desire our discrete theory to be as close as we can to the continuum one. 
We look at the inner product and what we can do with that next.

\subsection{The inner product is the star}
Any linear map on a vector space can be expressed in terms of the dual basis list
elements. If we have any $p$-form $\lambda$ we can wedge it to an $(n-p)$-form
$\mu$ to get an $n$-form $f d\sigma = \lambda \wedge \mu$. This is a linear map so $f$ 
is uniquely determined. Since it is linear we can also express it as $<\star\lambda,\mu>=f$
where $\star\lambda$ belongs to the dual space. This $\star$ is the hodge star and we can 
see how closely it is related to $<,>$. We cannot change one without changing the other.

We had in our original formulation that
\[ (e^i,e^j) = \delta_{ij} \]
which means that the inner product of an edge with itself is 1 whilst we'd
expect from the continuum case to get its length. This can be achieved by using 
a new inner product 
\[ (e^i,e^j) = \int W(e^i) \wedge \star W(e^j)\]
as used by Dodziuk \cite{Dodziuk}.
We want to investigate the effect this has on $\star$. We have introduced metric
information into our system and ideally this should filter through the system 
leaving all the properties which we are happy with whilst sorting out the metric
dependence of $\star$.

\subsubsection{Determining $\star$}
Our inner product is given by
\begin{equation}
 <e^i_B,e^j_B> = \int W^B(e^i_B)\wedge \star W^B(e^j_B),
\label{eqn1}
\end{equation}
Note that $e^i_B$ and $e^j_B$ both need to be of the same dimension
since we need an $n$-form on the $RHS$ for it to be nonzero. We then define
$\star_K$
as follows:
\begin{equation}
 <\star_K e^l_K,e^m_L> = \int W^B(e^l_K)\wedge W^B(e^m_L),
\label{eqn2}
\end{equation}
where using the fact that $e^l_K$ and $e^m_L$ can be expressed in terms of
elements of $B$.

Now we write
\begin{equation}
 \star_K e^l_K = h^l_m e^m_L.
\label{eqn3}
\end{equation}

We can using (\ref{eqn1}) and (\ref{eqn2}) determine $h^l_m$ which determines
our matrix
for the hodge star operator.

Firstly we define two more matrices $A$ and $B$ which allow us to move from
basis elements
of $K$ and $L$ to $B$ respectively. So
\begin{eqnarray}
 e^l_K &=& A^l_i e^i_B, \\
\label{eqn4}
 e^m_L &=& B^m_j e^j_B.
\label{eqn5}
\end{eqnarray}
and (\ref{eqn2}) can be rewritten as
\begin{eqnarray}
 <\star_Ke^l_K,e^m_L> &=& <h^l_o e^o_B,e^m_B>,\\
            &=& h^l_o B^o_i B^m_j <e^i_B,e^j_B>,\\
            &=& h^l_o B^o_i B^m_j \int W(e^i_B)\wedge \star W(e^j_B).\\
 I^{ij}     &=& \int W(e^i_B)\wedge \star W(e^j_B).\\
 <\star_Ke^l_K,e^m_L> &=& h^l_o B^o_i I^{ij} B^m_j,\\
            &=& h^l_o X^{om}.
           \label{eqnA}
\end{eqnarray}
Also using (\ref{eqn2}) we have that
\begin{eqnarray}
<\star_Ke^l_K,e^m_L> &=& \int W(e^l_K) \wedge W(e^m_L), \\
           &=& \int W(A^l_i e^i_B) \wedge W(B^m_j e^j_B), \\
           &=& A^l_i B^m_j \int W(e^i_B) \wedge W(e^j_B).\\
J^{ij}     &=&  \int W(e^i_B) \wedge W(e^j_B).\\
<\star_Ke^l_K,e^m_L> &=& A^l_i J^{ij} B^m_j ,\\
           &=& S^{lm}.
           \label{eqnB}
\end{eqnarray}

Since (\ref{eqnA})=(\ref{eqnB}) we know that 
\begin{equation}
h^l_o X^{om} = S^{lm},
\label{eqnstar}
\end{equation}
which is just a matrix equation. If we determine $X^{-1}$ we can right multiply
by this to get $h$ which is $\star_K$. We can get $\star_L$ using a similar calculation.

\begin{figure}[h]
\begin{center}
\includegraphics[width=150pt]{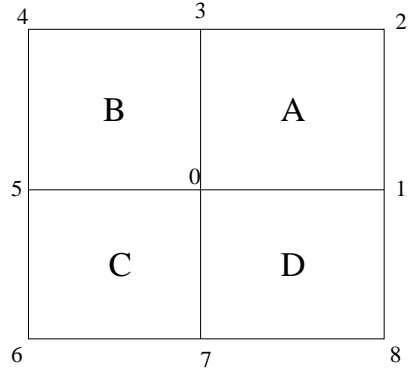}
\caption{The reference square.}
\end{center}
\end{figure}

\subsection{Cubic case}
Given an inner product, the Hodge star operator can be determined, as we 
have just seen. We consider here the cubic case\footnote{We have investigated the simplicial case
computationally but the nature of the volume factors is unclear due to its more complicated form. We
resulted with $\star\star$ diagonal but not equal to the identity; this could be normalised but
the metric dependence, which was the objective,  was not clarified.} and show that volume 
factors emerge in precisely the manner we expected from the heuristic.
We take a simple one square system which is subdivided as our reference system.   

We begin by determining the Whitney map of the various cells which we have.
For example 
\[
\begin{array}[t]{rlr}
W^B[0] &= \frac{1}{a^2}(a-x)(a-y) & A \\
     &=\frac{1}{a^2}(x)(a-y) & B \\
     &=\frac{1}{a^2}(x)(y) & C \\
     &=\frac{1}{a^2}(a-x)(y) & D \\
\end{array}
\]
\[
\begin{array}[t]{rlr}
W^B[7] &=-\frac{1}{a^2}(a-x)y & C \\
     &=-\frac{1}{a^2}(x)y & D \\
W^B[0123] &=\frac{1}{a^2}dx\wedge dy & A,B,C,D
\end{array}
\]

Once we have this we can look at $<,>_K$ where
\[ <x^p_K,y^p_K>_K=\int W^B(Bx^p_K)\wedge \star W^B(By^p_K).\]
If we let the $e^i$ be the basis list elements of $K$ then we can
define
\[ I^{ij}=<e^i,e^j>_K.\]
Note that
\begin{itemize}
\item I is diagonal otherwise $e^i$ and $e^j$ are in different squares of $B$ and
so their images under $W^B$ ( though not under $W^K$) do not overlap.
\item The various integrals that appear are the same since we are always integrating
either $x^2$ from $0$ to $a$ or something which can be expressed as this with a change
of variables.
\item The $dx$ and $dy$ integrals are independent and both of the form mentioned above.
\end{itemize}
The are only two possible cases which lead to the same result:
\begin{align*}
\int_0^a x^2 dx &=[\frac{x^3}{3}]_0^a &=\frac{a^3}{3}\\
\int_0^a (a-x)^2 dx &=\int_a^0 y^2 (-dy) &= \frac{a^3}{3}
\end{align*}
So any of the integrals which we have are equal to
$\frac{a^3}{3}$. 

In the case of vertices we have two such integrals
leading to $\frac{F\star A^6}{9}$ factors with a symmetry term $F$ to specify
how often a term occurs; this is 4 for vertices since each vertex occurs in 
4 squares, 2 for edges and 1 for faces. For edges we only have one integral, since
we don't get $x^2 y^2$ terms but $x^2dy$ ones instead, and for faces we
get no such terms since the integral is simply $dxdy$.  $B$ also introduces
a factor, as in the simplicial case; though here we get $2^p$ instead of $p!$
since the volumes involved are different\footnote{This is related to 
maintaining $A^KW^K=I$.}. Finally $W^K$ has
a $\frac{1}{a^2}$ term associated with it so for vertices we get
\[ <[0],[0]>_K=\frac{1 \times 4 a^6}{9 a^4}, \]
for edges
\[ <[01],[01]>_K=\frac{4\times 2 a^4}{3 a^4}\]
and for faces
\[ <[0123],[0123]>_K=\frac{16 \times 1 a^2}{a^4}. \]
For $L$ we similarly get
\begin{align*}
< \hat{[0123]},\hat{[0123]}>_L &=  \frac{1\times 4 a^2}{9} \\
< \hat{[01]},\hat{[01]}>_L &= \frac{4\times 2}{3} \\
< \hat{[0]},\hat{[0]} >_L &= \frac{16\times 1}{a^2} \\
\end{align*}
We only mention the results for particular vertices, 
edges and faces but they are all the same\footnote
{The $F$ factors are all the same since we are interested in complexes without
 boundary in which case every vertex has $F=4$, every edge has $F=2$ and 
every face has $F=1$.}. 

To determine $\star_K$ we use
\[ < \star_K x^p_K, y^{n-p}_L >_{L} = \int W^B(Bx^p_K)\wedge W^B(By^{n-p}_L). \]
We have the LHS except $\star_K$ so we need the RHS next.  Note that since  
both $<,>_K$ and $<,>_L$ are diagonal, we only get contributions when
$y=\hat{x}$.

We have two $B$ factors on the right hand side which leads to a 
factor of 4. If we are dealing with a vertex then we wedge it with a face.
 $B$ of the vertex gives 1, while $B$ of the face gives $4$. For edges we get 
$2\times 2$ since we wedge two edges together and for a face we wedge it with a
vertex leading to a $4\times 1=4$ factor. We look at the various cases next.

Performing the RHS integral for the vertices, if we have $x=[0]$ then
$y=\hat{[0]}$, we get:
\begin{align*}
 &=4\int_A \frac{1}{a^4}(a-x)(a-y) dx\wedge dy \\
 &=4 \frac{1}{a^4} \frac{a^2}{2} \frac{a^2}{2} \\
 &=1.
\end{align*}
For the edges we get:
\begin{align*}
&=4\int_A W^B[01] \wedge W^B[03] \\
&=4\int_A \frac{1}{a^4} \frac{a^2}{2}\frac{a^2}{2}\\
&=1.
\end{align*}
And for the the faces:
\begin{align*}
&=\int W^B[0123]\wedge W^B[0] \\
&=4\int_A \frac{1}{a^4} \frac{a^2}{2}\frac{a^2}{2}\\
&=1.
\end{align*}

The dual calculations are the same apart from sign factors which arise
when the $W^K(x)$ and $W^K(y)$ terms are flipped. So for the edges
case we get a $(-1)$ sign.

Determining $\star_K$ is now trivial since both sides of the
equation are diagonal. The $i$th diagonal element of $\star_K$ is simply the
$i$th diagonal element of the RHS(which are all just $\pm$ 1) divided by the
$i$th diagonal element of $<,>_L$ which only depends on whether dealing with
vertices, edges or faces. So on $0$-cochains have
\[ \star_K [0] = \frac{a^2}{16} \hat{[0]} \]
on edges we have
\[ \star_K [01] = \frac{3}{8}\hat{[01]} \]
and for faces we have
\[ \star_K [0123] = \frac{9}{4 a^2} \hat{[0123]}.\]

We can get $\star_L$ in the same way:
\begin{align*}
\star_L \hat{[0]} &=\frac{9}{ 4 a^2}[0] \\
\star_L \hat{[01]} &=-\frac{3}{8}[01] \\
\star_L \hat{[0123]} &=\frac{a^2}{16} [0123] \\
\end{align*}

With this we get
\[  \star_K \star_L = \pm\frac{9}{64}I \]
which is diagonal and proportional to the identity.
  
We can then introduce normalisation factors on the RHS  --- as is done in the 
simplicial theory --- by simply adding a factor of $\frac{8}{3}$ to the 
definition of $\star_K/\star_L$:
\begin{equation}
 < \star_K x^p_K, y^{n-p}_L >_{L} = \frac{8}{3}\int W^B(Bx^p_K)\wedge W^B(By^{n-p}_L). 
\label{eqn:norm}
\end{equation}
Using this we have $\star_K\star_L=\pm I$ as required. 

Here whilst we get $\star\star=I$ type behaviour, it can be improved. $\star_K$ acting on 
a vertex introduces a factor of $\frac{1}{6}$, after the above normalisation has 
been made. When $\star_L$ acts on this it multiplies it by $6$ so we get the desired
$\star\star=I$. Since does not occur in the continuum we should normalise at this level
instead which can be done by using appropriate $p$ dependent factors in Eq.\ref{eqn:norm}.

We can now determine convergence results since we can see how the length scale
$a$ appears in the various operators. In short they appear as you
would expect them to using the heuristic which means that
that $\delta$ and the Laplacian converge to the continuum case.

Firstly we look explicitly at what happens for $\delta=\star d\star$ in 2D.
In the continuum case we know that:
\begin{enumerate}
\item $\star d\star f = \star d fdxdy = 0 $
\item $\star d\star fdx = \star d (fdy) = \star\frac{\partial f}{\partial x} dx \wedge dy 
=\frac{\partial f}{\partial x}$
\item $\star d\star fdxdy = \star df = \star (\frac{\partial f}{\partial x}dx+
\frac{\partial f}{\partial y}dy)= \frac{\partial f}{\partial x}dy-
\frac{\partial f}{\partial y}dx.$
\end{enumerate}
We can now look at what happens in the cubic case and compare
$W^K \star_L d_L \star_K A^K$ with the above. 

\begin{figure}[h]
\begin{center}
\includegraphics[width=150pt]{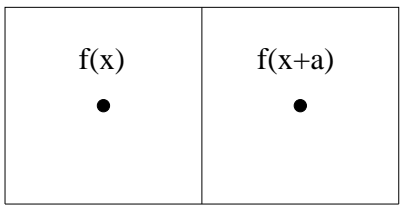}
\caption{Oppositely oriented adjacent regions lead to $f(x+a)-f(x)$ terms.}
\label{fig:pic}
\end{center}
\end{figure}

Note that $\partial=\star d\star$, with some sign factor which means that 
$\star d\star$ applied to $[xy]$ gives us its boundary, upto sign.
For example, if we take the edge $[0,y]$ we
map $-f_i$ onto it, where $f_i$ is $fdxdy$ integrated over the face 
$i$. The adjacent face will also contribute to this edge
with a $+f_j$ term. Face $i$ is $a$ away in the $x$ direction from 
face $j$ and so we effectively get an $f(x+a)-f(x)$ term which after
Whitney will give
\[ \frac{f(x+a)-f(x)}{a}dy. \]
In this way we can see how we agree with the continuum result. The $1$-form
case follows similarly with the boundary now vertices and the $0$-form 
case is trivial.

As we mentioned before, the volume factors which arise are precisely
 those which we required from our heuristic ( Sec.\ref{section:heuristic})
 which can be generalised to higher dimensions. 
First we note that this is the case in $2D$ since $\star$ applied to vertices leads to 
$a^2$ factor which is the volume of the cell mapped to divided by that of the cell 
acted on. For edges we get a factor independent of $a$ which again agrees with 
this picture as does the $\frac{1}{a^2}$ factor for faces.

In general we have $a^{3D-2p}/a^{2D}$ for $p$-cochains.
In two dimensions this means that we have $a^{6-2p}/a^4$ so $a^2$ for vertices, 
$1$ for edges and  $a^{-2}$ for faces. Similarly in three dimensions we have
$a^{9-2p}/a^{6}$ or $a^3$ for vertices, $a$ for edges, $a^{-1}$ for faces and
$a^{-3}$ for cubes. In general the
heuristic expression is the volume of the space mapper to, $a^{D-p}$, divided by the
volume of the object being acted on, $a^p$. Since 
\[
\frac{a^{3D-2p}}{a^{2D}}=a^{D-2p}=\frac{a^{D-p}}{a^p},
\]
the two agree.

We had already seen that the heuristic suggested a modification that could
be made to $\star^K$ so that it, along with the coderivative and Laplacian, would converge
to the continuum result. This problem with this was that it broke the relationship between
the inner product and $\star$. In order to retain this we made a natural modification 
to the discrete inner product to see what effect this had on $\star^K$ and   
found that this resulted in adding precisely the factors which 
we wanted from the heuristic. As a result, we can use the various convergence results 
which we had before, except that now the relationship of $\star^K$ and the inner product
is preserved.

\section{Conclusions}
We have thus introduced a cubic Whitney map which we have used to demonstrate convergence
of our discrete functions and operators to their continuum counterparts. This involved
the introduction of some modifications for the Hodge star, either via a heuristic, which
was unsatisfactory due to the relationship of the inner product and $\star$, or
via a new inner product (the one in fact used by Dodziuk).

With metric and convergence results, applications become a possibility. The work,
for example, of Bossavit \cite{Bossavit}, Hiptmair \cite{Hiptmair} and Nedelec \cite{Nedelec} 
are naturally of interest though within a different context, namely finite element. GD also involves 
complexes, with finite dimensional subpaces within them, but does not use variational arguments 
for existence theorems for example.  

Our current focus is on application to lattice field theory where following Rabin \cite{Rabin}
we have shown that within the Dirac-K\"ahler formalism \cite{BJ}, where Nielsen-Ninomya \cite{NN} is 
not applicable as shown by Becher \cite{Becher}, we have chirality ($\star$ in this
picture) whilst avoiding degeneracy. 

Work has also been done extending the mathematical structures which GD deals and should
soon follow. 
\section{Acknowledgments}
Thanks to many people without whom this work would not have been possible. David Adams
for developing the core work and giving me something to do, James Sexton for telling me about
the link with Dirac-K\"ahler as well as supervising me and of course my dad.


\begin{thebibliography}{99}
\bibitem{Adams}
D.~H.~Adams, Phys. Rev. Lett. {\bf 78}, 4155 (1997) ; T.C.D. Preprint
hep-th/9612009.
\bibitem{Us}
Samik Sen, S.~Sen, J.~C.~Sexton and D.~H.~Adams, 
Phys. Rev. {\bf E61}, 3174-3185 (2000).
\bibitem{me}
Samik Sen, {\em Extending Geometric Discretisation}, Ph.D. Thesis, University of Dublin (2001).
\bibitem{Dodziuk}
J.~Dodziuk, {\em Finite-Difference approach to the Hodge Theory of Harmonic
Forms}, American Journal of Mathematics, Vol. 98, No. 1, 79-104 (1976).
\bibitem{Hiptmair}
R.~Hiptmair, {\em Discrete Hodge operators:An algebraic perspective}, Geometric Methods for Computational
Electromagnetics, F. Teixeir, ed, vol. 32 or PIER, EMW Publishing, Cambridge, MA, 2001, pp247-269.
\bibitem{Nedelec}
J. Nedelec,{\em Mixed finite elements in $R^3$}, Numer. Math., 35  314-341 (1980).
\bibitem{fermions}
V. de Beauce and Samik Sen, {\em Chiral Fermions on the Lattice}, hep-0305125.
\bibitem{Rabin}
J.~M.~Rabin, Nuclear Physics {\bf B201}, 315-322 (1982).
\bibitem{Bossavit}
A.~Bossavit, {\em Computational Electromagnetism}, Academic Press,
London (1998).
\bibitem{Nakahara}
M. Nakahara, {\em Geometry, Topology and Physics}, Institute of Physics Publishing  
(1990).
\bibitem{NS}
C.~Nash and S.~Sen, {\em Topology and Geometry for Physicists}, Academic
Press, London (1983).
\bibitem{Eguchi}
T.~Eguchi, P.~B.~Gilkey and A.~J.~Hanson, {\em Physics Reports} {\bf 66},
 No.6 , 213-393, North-Holland Publishing Company (1980).
\bibitem{Rado}
T. Rad\'o, \"Uber den Begriff der Riemannschen Fl\"ache, 
{\em Acta Litt.~Sci.~Szeged.} {\bf 2} (1925);
Proof in Chapter 1 of {\em Riemann Surfaces}, Princeton Unversity
Press, 1960 by L.~V.~Ahlfors and L.~Sario.
\bibitem{HY}
J.~G.~Hocking and G.~S.~Young, {\em Topology}, Dover Publications, NY (1961). 
\bibitem{Whitney}
H.~Whitney, {\em Geometric Integration Theory}, Princeton University Press,
Princeton, NJ, (1957).
\bibitem{starproduct}
D.~Birmingham, M.~Rakowski, Phys. Letters {\bf B299}, 299 (1993).
\bibitem{Albeverio}
S.~Albeverio and J.~Sch\"{a}fer, J. Math. Phys. {\bf 36}, 2157 (1995).
\bibitem{BJ}
P.~Becher and H.~Joos, Z.~Phys. {\bf C15}, 343 (1982).
\bibitem{NN}
H.~B.~Nielsen and  M.~Ninomiya, Nuclear Physics {\bf B185}, 20-40 (1981).
\bibitem{Becher}
P.~Becher, {\em Phys. Lett.}, {\bf 104B}, 221 (1981).
\end{thebibliography}
\end{document}